\documentclass[aps,prl,twocolumn,superscriptaddress,showpacs]{revtex4}
\usepackage{graphicx}
\usepackage[floatfix]{epsfig}
\usepackage{dcolumn}
\usepackage{bm}
\usepackage{color} 

\def\tc{T$_c$ }
\def\sc{superconducting }
\def\af{antiferromagnetic }

\def\beq{\begin{equation}}
\def\eeq{\end{equation}}
\def\eqref#1{(\ref{#1}) }
\def\sof{\mbox{$SO(5)$} }
\long\def\joesel#1{{ #1}}
\newcommand{\helic} {\Upsilon}
\def\mathit#1{#1}

\newcommand{\hc}    {\mbox{h.c.} }
\newcommand{\op}[1] {\mbox{$\hat{#1}$}}
\def\domybib{1}
\begin{document}

\title{Scaling properties of the projected \sof model in three dimensions}

\author{M. J\"ostingmeier}
\affiliation{Institut f\"ur Theoretische Physik, Universit\"at
  W\"urzburg, D-97074 W\"urzburg, Germany}
\author{A. Dorneich}
\affiliation{Institut f\"ur Theoretische Physik, Universit\"at
  W\"urzburg, D-97074 W\"urzburg, Germany}
\author{E. Arrigoni}
\affiliation{Institut f\"ur Theoretische Physik, Universit\"at
  W\"urzburg, D-97074 W\"urzburg, Germany}
\affiliation{Department of Physics, Stanford University,  Stanford,
  94305 California, USA}

\author{W. Hanke}
\affiliation{Institut f\"ur Theoretische Physik, Universit\"at
  W\"urzburg, D-97074 W\"urzburg, Germany}
\author{Shou-Cheng Zhang}
\affiliation{Department of Physics, Stanford University, Stanford, 94305 California}

\date{\today}

\begin{abstract}
We study the scaling properties of the quantum ``projected'' \sof model in three
dimensions by means of a highly accurate Quantum-Monte-Carlo analysis.
Within the parameter regime studied (temperature and system size),
we show that the scaling behavior is consistent with a
$\sof$-symmetric critical behavior in a relative extendent transient regime.
This holds both when the symmetry breaking is caused by quantum
fluctuations only as well as  when also the static (mean-field)
symmetry is moderately broken.
We argue that possible departure away
from the \sof-symmetric scaling occurs only in an extremely narrow parameter
regime, which is inaccessible both experimentally and numerically.
\end{abstract}

\maketitle

\section{Introduction}

A common feature of the phase diagram of most high-\tc superconductors
(HTSC) is the close proximity of the \sc (SC) and the \af (AF) phases.
The \sof theory of high-\tc SC describes the transition between these
two phases by an effective quantum non-linear $\sigma$ model with
approximate $\sof$ symmetry, which unifies the two order
parameters~\cite{zhan.97}.
Several microscopic \sof-symmetric models have been proposed which
succesfully describe many features of the cuprate
physics~\cite{ed.do.98,de.ko.98,hu.01,ar.be.97,so5.97}.
A breakthrough has been recently achieved  by
a model that reconciles the \sof symmetry with the physics of the Hubbard
gap~\cite{zh.hu.99,za.ha.00}.
In the so-called ``projected'' \sof (p\sof) model, the Gutzwiller
projection is implemented exactly.
Although the {\it exact }  symmetry is destroyed at the
microscopic level due to quantum
fluctuations originating from the projection,
it would be interesting to investigate whether an effective \sof 
symmetry
 still controls the
asymptotic behavior of the system 
 at long-wavelength, i. e., \sof symmetry is asymptotically restored.
This has been shown to be the case,
 for example, for 
a two-leg ladder~\cite{so5.97} system. 
This behavior is physically similar to
the case of interacting Fermi liquids in which
the characteristic symmetries of the non-interacting system -- namely
 conservation of the particle number for each momentum $k$ of the
 Fermi surface -- are
 destroyed by the interaction, but are asymptotically restored
for low-energy excitations close to the Fermi surface.
In contrast to the ladder, for a higher-dimensional system,
the candidate for symmetry restoration should be sought at finite
temperatures
and is provided by the multicritical point  where the AF and SC
critical lines meet.

The problem of symmetry restoration at this multicritical point
 has been addressed  by Hu in Ref.
\onlinecite{hu.01} via a Monte Carlo (MC) calculation of a classical
\sof model in which an additional quartic anisotropy term has been included.
Classical MC are by orders of magnitude easier to perform
and less resource demanding than QMC simulations, hence very large
system sizes can be simulated and highly accurate data are obtained.
For this reason Hu could carry out a detailed analysis of the AF-SC phase
diagram and of the critical behavior.
In particular, he could show that the SC helicity modulus follows a
scaling law determined by the corresponding \sof critical exponent
$\nu_5$ and by the crossover exponent $\phi$. For these exponents, Hu
could find a good agreement with the results of the
$\epsilon$-expansion.
On the other hand it was pointed out by
Aharony~\cite{ahar.02.comm}
via a rigorous argument that the decoupled fixed point is stable, and he further
concluded 
 that neither the biconical nor the $\sof$ fixed points are stable.
However, he also commented that the unstable flow is extremely slow for the $
\sof$ case due to the small crossover exponent.

In this paper, we address this issue by a detailed numerical analysis of the
phase diagram and of the
critical properties of the three-dimensional p\sof model. Our strategy is, thus, to
start from a  model in which \sof symmetry is {\it realistically}
broken and to investigate to what extent this \joesel{symmetry} 
is restored at the
multicritical point.
Our main results are twofold. Within the system sizes and temperature
ranges we can achieve, the scaling behavior is consistent with an
\sof critical behavior. 
On the other hand, since the \sof
fixed point could be ultimately unstable~\cite{ahar.02.comm},
 we make an analytical estimate -- based on
the $\epsilon$-expansion -- of the size
of the critical region in which deviations from the \sof behavior
should be observed. Due to the small crossover
exponent~\cite{ahar.02.comm}
 it turns out
that the unstable flow only takes effect when
the reduced temperature measured from the bi-critical point is 
very small.
 Therefore, the possible unstable effect
can neither be observed experimentally nor numerically. For all practical
purposes, the multicritical point is dominated by the initial flow
towards the $\sof$-symmetric behavior\cite{mu.na.00}.
Again,
this situation is very similar to the case of  Fermi liquids.
Due to the well-known Kohn-Luttinger effect~\cite{ko.lu.65}, the Fermi-liquid fixed
point is always unstable towards a SC state.
However,
 this effect is experimentally
irrelevant for most metals since it only works  at exponentially low temperatures.

\section{The model}

In the p\sof
model each coarse-grained lattice site represents a plaquette of the
original lattice model, and the lowest energy state on the plaquette
is a spin singlet state at half-filling.
 There are four types
of excitations, namely, three magnon modes
 and a hole-pair mode.
\joesel{Their dynamics are described by the following Hamiltonian:
\begin{eqnarray}
\op{H} &=&
\Delta_s \!\!\!\!\!\sum_{x,\alpha{\scriptscriptstyle =2,3,4}}
  \!\!\!\!t_\alpha^\dagger{\scriptstyle (x)}t_\alpha{\scriptstyle (x)}
  + (\Delta_c\!- 2\mu) \sum_{x} t_h^\dagger{\scriptstyle(x)}
     t_h{\scriptstyle (x)}\\
  &-& J_s\!\!\!\!\!\!\!\!\!\!
  \sum_{ <xx'>,\alpha{\scriptscriptstyle =2,3,4}}
        \!\!\!\!\!\!\!\! n_\alpha{\scriptstyle (x)} n_\alpha{\scriptstyle (x')}
  -\!J_c\!\!\! \sum_{ <xx'> }\!\! (t_h^{\dagger}{\scriptstyle (x)}
    t_h{\scriptstyle (x')} \!+\! \hc)\,,\nonumber
\label{H_pSO5}
\end{eqnarray}
Here $t^{(\dagger)}_{\alpha=2,3,4}$ anihilates (creates) a triplet
state, $t^{(\dagger)}_h$ anihilates (creates) a hole pair state and
\mbox{$n_\alpha=(t_\alpha + t^\dagger_\alpha)/\sqrt{2}$}
are the three components of the N\'eel order parameter.
$\Delta_s$ and  $\Delta_c\sim U$ are
the energies to create a magnon and a hole-pair excitation,
respectively, at vanishing chemical potential $\mu=0$.
This model can also be effectively obtained by a coarse-grained
reduction
of more common models such as $t-J$ or Hubbard~\cite{al.au.02}.
In order to study the effect of symmetry breaking we
  consider different situations associated with different 
sets of parameters. First, we consider the case where
$J_s=J=J_c/2$ (our zero of the chemical potential is such that $\Delta_s=\Delta_c$).
It has been shown~\cite{zh.hu.99} that this model
 has a static \sof symmetry at
the mean-field level and that the symmetry is only broken by quantum
fluctuations~\cite{ar.ha.00}.
Since we want to carry out our analysis also
for a more realistic
  model in which also the static
\sof\ symmetry is broken,
 we also consider a system with a different  ratio $J_s/J_c$. 
In particular, one would like to 
reproduce the order of mangitude of $T_c/T_N$ observerved in the
cuprates, where $T_c$ $(T_N)$ denominates the SC critical temperature
(N\'eel Temperature).
However, this behavior is obtained for 
$J_s/J_c\sim 2$, for which the numerical simulation is rather unstable,
making it impossible to determine the critical exponents with
sufficient accuracy.
For this reason, we choose a value of the parameter ``in between''
$(J_c=J_h=J)$, for which also the static \sof\ symmetry is broken.}

The phase diagram of
this model in two dimensions has been analyzed in detail by a
numerical Quantum-Monte-Carlo approach in
Ref.~\cite{do.ha.02}. In particular, the model has been shown to
provide a
semiquantitative description of many
properties of the HTSC in a consistent way.
In Ref.~\cite{do.ha.02}, the SC transition has been identified as
 a Kosterlitz-Thouless phase in which the SC correlations decay
 algebraically.
Unfortunately, there is no such transition for the AF phase in
two dimensions, as
all AF correlations decay exponentially at finite temperatures.
Therefore, in order to analyze the multicritical point where the AF and SC
critical lines meet, it is necessary to work in three dimensions,
which is what we investigate in the present paper.

\section{Results}

\subsection{ Case $J_s=J_c/2$ }

\begin{figure}[htb]
\begin{center}
\epsfig{file=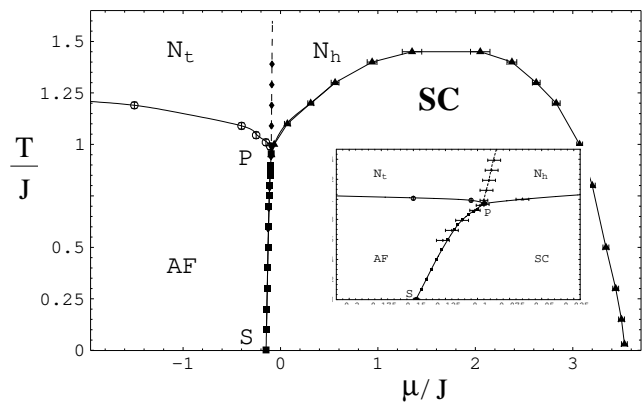,width=\columnwidth}
\end{center}
\caption{
\label{fig_pd_mu}
Phase  diagram $T(\mu)$ of the three-dimensional p\sof model
with $J=J_s\!=\!J_c/2$ and $\Delta_s\!=\!\Delta_c\!=\!J$.
N$_h$ and N$_t$ are, respectively,  the hole-pair and the magnon-dominated
regions
of the disordered  phase. The separation
line between N$_h$ and N$_t$ is the line of equal spatial correlation
decay of hole-pairs and bosons.
The inset shows a detailed view of the region near the multicritical point.
}
\end{figure}

\begin{figure}[htb]
\begin{center}
\epsfig{file=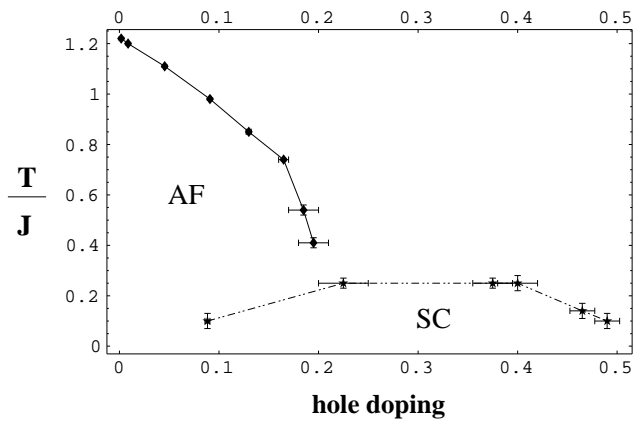,width=\columnwidth}
\end{center}
\caption{
\label{jratio}
Phase diagram for
$J_c/(2 J_s)=0.225$ as function of the hole doping $\delta$.
}
\end{figure}

We start by presenting
the phase diagram of the 3D
p\sof model for the ``symmetric''  case $J_s=J_c/2$.
 Figure \ref{fig_pd_mu} shows an AF and a
SC phase extending to
 finite temperatures as expected.
Furthermore, the two phase transition lines
merge into a multicritical  point (at $T_b\!=0.960\pm 0.005$
and $\mu_b\!=\!-0.098 \pm 0.001$).
The line of equal correlation decay of hole-pairs and triplet bosons
also merges into this multicritical point $P$.
Unlike the corresponding phase
in the classical model,
the SC phase extends only over a finite $\mu$ range; this is
due to the hardcore constraint of the hole-pair bosons
and agrees with experimentally determined phase diagrams of the
cuprates. In this \joesel{sense}, the quantum mechanical
p\sof model is 
more physical  than the classical \sof model.
However, in real cuprates the ratio between the maximum SC temperature $T_c$
and N\'eel temperature
$T_N$ is about 0.17 to 0.25,
whereas in the p\sof model we obtain the values
$T_c/J\!=\!1.465\pm 0.008$ at $\mu_{\mathit{opt}}/J\!\approx\!1.7$
and $T_N/J\!=1.29\pm 0.01$,
hence $T_c$ is slightly larger than
$T_N$.
In order to obtain realistic values for the transition temperatures,
it is necessary to relax the {\it static} \sof condition
and take a smaller value for the ratio
$J_c/(2 J_s)$, \joesel{which breaks \sof symmetry even on a
  mean field level.}  
The phase diagram with
$J_c/(2 J_s)=0.225$ is plotted
in Fig.~\ref{jratio}. As one can see, this gives a more realistic
ratio of $T_N/T_c\approx 0.2$.
However, it should be pointed out that the numerical effort
to treat such different values of $J$ is order
of magnitudes larger than considering 
 $J_c$
and $J_s$ of the same order of magnitude, as we have done in Fig.~\ref{fig_pd_mu}.
 Therefore, 
we will also consider a system 
with $J_c= J_s=1$ 
for which also the static \sof symmetry
is broken.
For the same reason, we neglect here the $c$-axis anisotropy and consider an
isotropic 3D model.

We first carry out an analysis of the critical properties for
$J_c/(2 J_s)=1$
A closer look to the phase transition line between the points $S$ and $P$
reveals (inset of Fig.~\ref{fig_pd_mu}) that this line
is not vertical as in the classical \sof model but slightly inclined.
This indicates that a finite latent heat is connected with the AF-SC
phase transition. Moreover, this means that in contrast to the classical
model, $\mu$ is not a scaling variable for the bicritical point $P$.

\subsection{Scaling analysis}

We now perform a scaling analysis similar to the one performed
by Hu~\cite{hu.01} in a classical \sof system. The most important
outcome of this analysis will be the strong numerical
\joesel{indication} that in a large region around the multicritical
point  the full \sof symmetry is
approximately restored. This  is non trivial  for a system whose
\sof-symmetry has manifestly been broken by projecting out all
doubly-occupied states.
First we want to determine the form of the $T_N(\mu)$ and $T_c(\mu)$
curves in the vicinity of the bicritical point.
For crossover behavior 
with an exponent $\phi>1$
one would generally expect
 that the two curves merge tangentially into
the first-order line. However, this holds
for the scaling variables,
therefore, one should first
 perform a transformation from the old
$\mu$ axis to a new $\mu'$ axis defined by
$  \mu'(T) = \mu - (T-T_b)/m\,,$
where $m\approx 0.11$ is the slope of the first order line below $T_b$.

\begin{figure*}
\epsfig{file=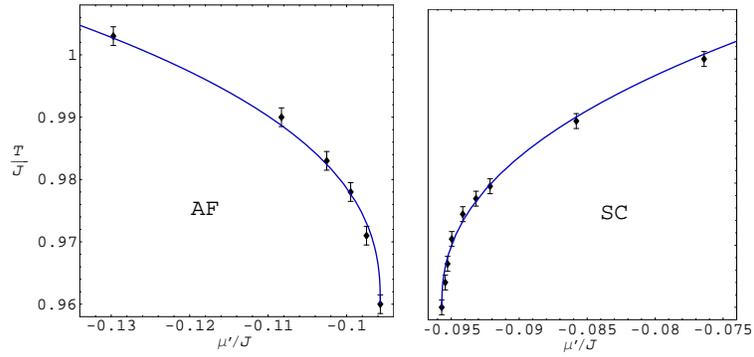,width=10cm}
\caption{
\label{fig_pSO5_AFB3_SCB2}
Plot of the AF (left) and SC (right) critical lines in the vicinity of the
multicritical point.
}
\end{figure*}

After this transformation,
the
transition curves
 $T_N(\mu')$ and $T_c(\mu')$
are quite well described by the crossover behavior
(we now drop the prime for convenience)
 \begin{eqnarray}
 &&  \frac{T_c(\mu)}{T_b}-1 = B_2\cdot (\mu-\mu_b)^{1/\phi}
\nonumber \\
 \label{eq_B2_B3}
 \mbox{and}\quad\; &&
 \frac{T_N(\mu)}{T_b}-1 =  B_3\cdot (\mu_b-\mu)^{1/\phi}
 \end{eqnarray}
 The  fit to this behavior is shown
in more detail in Fig.\ \ref{fig_pSO5_AFB3_SCB2}. However,
the value of $\phi$ we obtain
 ($\phi\approx 2.35$)  is considerably larger
than the value expected form the $\epsilon$-expansion.
It should be noted that the
above determination of $\phi$ is not very accurate: the data points in
Fig.\ \ref{fig_pSO5_AFB3_SCB2} are the result of a delicate finite-size
scaling for lattice sites up to $18^3$,
followed by the transformation from $\mu$ to $\mu'$ which again
increases the numerical error bars.
For this reason it cannot be excluded that
the difference in the $\phi$ values is mainly due to statistical
and finite-size scaling errors.
In fact, a more accurate evaluation of $\phi$ will be provided below.

 On the
SC side, the finite-size scaling
carried out in order to extract the order parameter and the transition
temperature
turns out to be quite reliable. On the other hand,
on the AF side, the fluctuations in the particle numbers of the
three triplet bosons slightly increase the statistical errors of the SSE
results and make the finite-size scaling more difficult.

The critical exponents for the onset of AF and SC order as a function of
temperature for various chemical potentials can be extracted from
Fig.\ \ref{fig_pSO5_AFB3_SCB2}.
Far into the SC range, at $\mu\!=\!1.5$,
we find for the SC helicity modulus~\cite{fi.ba.73}
\[
  \helic \propto  (1-T/T_c)^{\nu}\quad\;\mbox{with}\quad\;\nu=0.66\pm 0.02\,,
\]
which matches very well the values
obtained by the $\epsilon$-expansion and by numerical analyses of
a 3D XY model.
On the AF side, error bars are larger, as discussed above. We obtain
for the AF order parameter
\[
{C_{AF}(\infty)}
 \propto  (1-T/T_c)^{\beta_3} \quad\; \mbox{with}\quad\;\beta_3 = 0.35 \pm 0.03,
\]
for $\mu=-2.25$,
also in accordance with the value expected for a 3D classical
Heisenberg model.

In order to determine $\nu$ and $\phi$ more accurately in the crossover regime,
we use two expressions derived from the scaling behavior (cf. Ref.~\cite{hu.01}).
  \begin{equation}
  \label{eq_nu5_div_phi}
  \helic(T_b,\mu) / \helic(T_b,\mu'')
   = \big( (\mu-\mu_b)/(\mu''-\mu_b) \big)^{\nu_5/\phi}\,.
  \end{equation}
and
  \begin{equation}
  \label{eq_phi}
  \phi = \frac{
\ln\Big(\frac{\mu_2-\mu_b}{\mu_1-\mu_b} \Big)
           }{
    \ln\bigg( \frac{\partial}{\partial T}\frac{\helic(T,\mu_1)}
    {\helic(T,\mu_1')} \Big|_{T=T_b} \Big/
    \frac{\partial}{\partial T}\frac{\helic(T,\mu_2)}
    {\helic(T,\mu_2')} \Big|_{T=T_b} \bigg)
}
\,
  \end{equation}
  where $\mu_1$, $\mu_1'$, $\mu_2$, and $\mu_2'$ are related by
  $(\mu_1-\mu_b)/(\mu_1'-\mu_b) \!=\! (\mu_2-\mu_b)/(\mu_2'-\mu_b) > 0$.

\begin{figure}[htb]
\epsfig{file=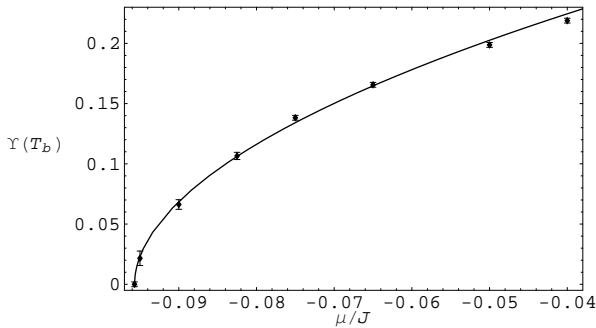,width=\columnwidth}
\caption{
\label{fig_pSO5_nu5phi}
Helicity $\helic$ as a function of the chemical potential $\mu$ at
$T\!=\!T_b$. From this function, the value of $\nu_5/\phi$ can be extracted
via equation (\ref{eq_nu5_div_phi}).}
\end{figure}

The result is shown in Fig.\ \ref{fig_pSO5_nu5phi}:
we obtain the ratio
\[
  \nu_5/\phi = 0.52 \pm 0.01,
\]
which is in excellent accordance with
the results of the
$\epsilon$-expansion and other numerical analyses~\cite{hu.01}.
$\phi$ is then obtained by using
\eqref{eq_phi}.
We have applied \eqref{eq_phi} onto 9
different combinations of $(\mu_1,\mu_1'\!=\!\mu_2,\mu_2')$ values
with $\mu_1/\mu_1'\!=\!\mu_2/\mu_2'\!=\!0.5$.
The result is
\[
  \phi = 1.43 \pm 0.05\,,
\]
which is again in good agreement with
the $\epsilon$-expansion for a \sof bicritical point and with the
results of Ref.~\cite{hu.01}.

\subsection{ Case $J_s=J_c$ }

\joesel{This  agreement between the critical exponents obtained in the
  previsous section may not come
completely as a
surprise, since the \sof symmetry is only broken by quantum
fluctuations for the parameter we have taken.
 The question we want to adress now is wether \sof
  symmetry is also asymptotically restored for 
a more realistic set of parameters for which the static \sof symmetry is 
broken as well.
 As already mentioned above, 
the case, where the phase diagram of the cuprates is qualitatively
well reproduced ($J_c/(2J_s)= 0.225$, see Fig.\ \ref{jratio}), is too
difficult to 
address numerically, so that the critical exponents cannot be
determined with sufficient precision in this case.
Therefore, we
repeat our analysis
 for the model in an intermediate regime ($J_c=J_h$), which is not so
 realistic but for which the static \sof symmetry is  broken as well.
One could hope that if \sof symmetry is restored for here,
then it might be also restored
 for the case   $J_c/(2J_s)= 0.225$, although one may expect
 that the asymptotic region in which this occurs
 will be less extended.
We stress again the fact that eventually one should expect the system to
flow away from the \sof fixed point, although in a very small critical
region~\cite{ahar.02.comm}.}

\begin{figure}[htb]
  \epsfig{file=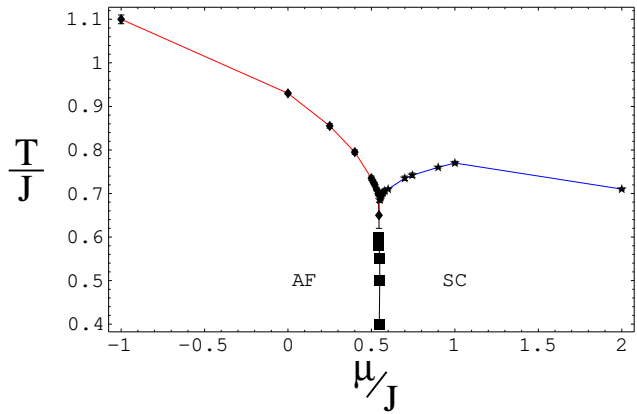,width=\columnwidth}
  \caption{
    \label{jratio2}
    Phase diagram as a function of the chemical potential for
    $J_c=J_h=1$, the lines are guides to the eyes.}
\end{figure}

\begin{figure}[htb]
  \epsfig{file=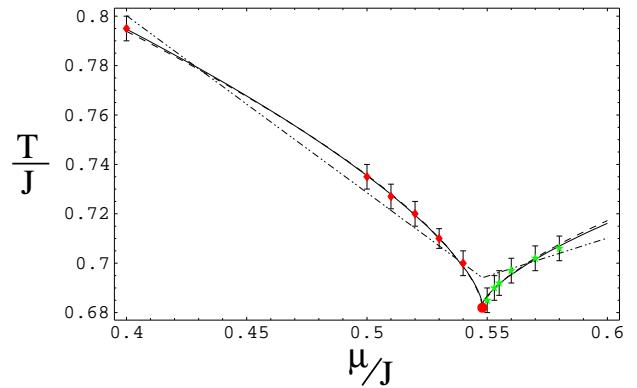,width=\columnwidth}
  \caption{
    \label{fitted_phdiag}
    Detailed view of the phase diagram as a function $\mu$ ($J_c=J_h=1$),
    the two lines have been obtained by fits to  
    \mbox{$T(\mu)=T_b*\left(1 + (B_2 + B_3*Sign[\mu_b - \mu])*
        \mid \mu-\mu_b]^{\frac{1}{\phi}} \right)$.} The continuous
    (dashed) line is the `normal' (`weighted') fit. 
    The decoupled fixpoint case is plotted as a dashed-dotted line.}
\end{figure}

\begin{figure}[htb]
  \epsfig{file=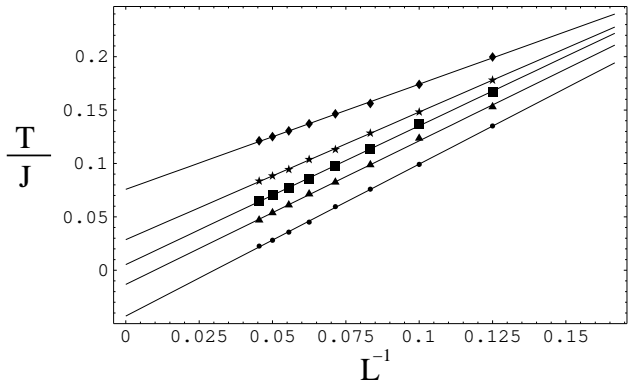,width=\columnwidth}
  \caption{
    \label{finite_size}
    Finite size scaling of the antiferromagnetic order parameter for
    $\mu=0.5$, the temperatures cover 0.72J (lozenge), 0.73J (star),
    0.735 (square), 0.74J (triangle) and 0.75 (cirle).
    The lattice size was varied from 216 
    $(8^3)$ upto 10648 $(22^3)$ sites, scanning all cubes with even
    edge length.}
\end{figure}

\joesel{The  phase
  diagram for $J_c=J_h$
is presented in Fig. \ref{jratio2} and a detailed view of the
  region close to the bicritical point is plotted in Fig. \ref{fitted_phdiag}.
  Here, the points in the plots were obtained by a finite-size scaling
  with lattices up to 5032 $(18^3)$ sites. In some cases, we were able
  to simulate
  lattices up to 10648 $(22^3)$ sites. 
An example of the finite-size scaling is shown in Fig. \ref{finite_size}. 
Our analysis yields
$T_b = 0.682 \pm 0.005$ and $\mu_b = 0.548 \pm 0.0005$.
  Here the line of equal correlation decay is vertical within the
  error bars, so the
  transformation from $\mu$ to $\mu^\prime$ is not necessary and
  the error bars are not increased by the transformation.
  This allows to determine the critical exponents by fitting the data
  points visible in Fig. \ref{fitted_phdiag}
  against
  \mbox{$T(\mu)=T_b*\left(1 + (B_2 + B_3*Sign[\mu_b - \mu])*\mid x -
      \mu_b\mid ^\frac{1}{\phi}\right)$.} 
  We obtain:
  \begin{eqnarray}
    B_2&=& 0.47 \pm 0.07,\label{B2} \\ 
    B_3&=& 0.11 \pm 0.04,,\label{B3}\\
    \phi&=& 1.49 \pm 0.18,\label{phinormal}\\
    Tb&=&0.683\pm 0.004,\label{Tbnormal}\\
    \frac{B_2}{B_3}&=&1.67, \pm 0.36\label{B2oB3normal}
  \end{eqnarray}
Since points further away from the bicritical point are expected to
show a larger deviation  from
the bicritical behavior, 
  we also performed a weighted fit, which takes this fact into
  account.
 Here, data points closer to the
  bicritical point are weighted more than the ones further away.
Specifically, in both the SC and the AF phase,
the point closest to the bicritical point is
  weighted six times  the one with the largest distance to the
  bicritical point. The second closest is weighted 5 times and so
  on. The results are, within the error bars, quite similiar to the
  ones obtained without this different weighting procedure: 
  \begin{eqnarray}
    B_2&=& 0.46 \pm 0.05,\\ \label{B2weighted}
    B_3&=& 0.11 \pm 0.03,\label{B3weighted}\\
    \phi&=& 1.53 \pm 0.12\label{phiweighted}\\
    Tb&=&0.682\pm 0.003\label{Tbweighted}\\
    \frac{B_2}{B_3}&=&1.61 \pm 0.23\label{B2oB3weighted}
  \end{eqnarray}
The agreement between Eqs.~\ref{B2}-\ref{B2oB3normal} and
Eqs.~\ref{B2weighted}-\ref{B2oB3weighted} suggests that
the data we have considered are still controlled by the bicritical point,

In order to test whether alternativly proposed fixed points may be excluded, we 
carried out a least-square fit
of
 our data to
  the decoupled fixpoint case ($\phi=1, B_2, B_3$ and $T_b$
  arbitrary). 
This is shown in Fig. ~\ref{fitted_phdiag} (dashed-dotted line).
As one can see form the curve,
our data doe not support
  the assumption that the 
  p\sof model has a decoupled fixpoint.
Since we know that ultimately the decoupled fixed point should be the
stable one~\cite{ahar.02.comm}, this means that our systems is
still quite far
away from the ultimate asymptotic region.
}

\section{Discussion and conclusions}

Altogether, the scaling analysis of the 3D p\sof model has produced
a crossover exponent which matches quite
well with the corresponding value obtained from a classical \sof
model and from the $\epsilon$-expansion.
This gives convincing evidence that the static correlation functions
at the p\sof multicritical point is controlled by a
fully \sof symmetric point, at least in a large transient region.
However, one should point out that within the statistical and
finite-size error, as well as within the error due to the
extrapolation of the $\epsilon$-expansion value to $\epsilon=1$ one
cannot exclude that the actual fixed point one approaches is the
biconical one, which has very similar exponents to the isotropic \sof
one.
On the other hand, the biconical fixed point should be accompanied by a
AF+SC coexistence region (as a function of chemical potential), which
we do not observe.
As discussed above we can certainly exclude in this transient region
the {\it decoupled} fixed
point for which 
$\phi=1$\cite{ahar.02.comm}.
Of course, our limited system sizes cannot tell which fixed point
would be ultimately stable 
in the deep asymptotic region.
Here, Aharony's exact statement shows that the decoupled fixed point
should be ultimately the stable one in the deep asymptotic region~\cite{ahar.02.comm}.

We argue 
that the resolution between this exact result
 and the numerically
observed \sof critical behavior lies in the size of the critical
region~\cite{ahar.02.comm}. 
We now give an estimate, based on $\epsilon$ expansion, for
 the scale at which the instability of the \sof
fixed point could be detectable.
This estimate holds for the case in which one has a ``static''
\sof symmetry at
the mean-field level.
The symmetry-breaking effects due to quantum fluctuations have been
estimated in Ref.~\cite{ar.ha.00} and are given by Eq.~(36) there.
By replacing the initial conditions for the bare couplings in terms of
the microscopic parameters of the Hamiltonian
(cf. Eq. 26 of Ref. ~\cite{ar.ha.00}), and projecting along the
different scaling variables around the \sof fixed point, one obtains
a quite small projection
along the variable that scales away from the
fixed point. Combined with the fact that the exponent for this scaling
variables is quite small 
($\lambda = 1/13$ at the lowest-order in the $\epsilon$ expansion, 
although more accurate estimates~\cite{ca.pe.02u,ca.pe.03,pe.vi.02}
give a somewhat larger value of $\lambda \approx 0.3$),
 we obtain an estimate for
the scaling region in which the \sof fixed point is replaced by
another -- {\it e.g.} the biconical or the decoupled -- fixed point
at $t\equiv (T_b-T)/T_b \sim 10^{-10}$ if one takes the $O(\epsilon)$ result
for the exponent.
Notice that taking the result of Ref.~\cite{ca.pe.03} for the exponent, one
obtains a quite larger value
$t\sim 2. 10^{-3}$. However, since
                  the multi-critical temperatures 
of relevant materials (organic conductors, and, more recently, $YBa_2Cu_3O_{6.35}$) 
are around $10 \ K$, the critical region is still basically unaccessible experimentally 
as well as with our {\it quantum} simulation. 
On the other hand, the other scaling variables, although being
initially of the order of $1$, rapidly scale to zero due to the large,
negative, exponents. Therefore, the  \sof regime starts to become
important as soon as the AF and SC correlation lengths become large and
continues to affect the scaling behavior of the system  basically in
the whole  accessible region.

In conclusion, we have shown \joesel{numerically}  that the p\sof model, which
combines the 
idea of \sof symmetry with a {\it realistic} treatment of the Hubbard
gap,\joesel{ reproduces the salient features of the cuprate's phase diagram.
Furthermore, this model is controlled by a \sof symmetric bicritical
point, at least 
within a large transient region.
We have shown that this also holds
for a case in which the static
\sof symmetry is broken.} 
Possible flow away from the symmetric
fix point occurs only within an extremely narrow region in reduced
temperature, making it impossible to observe both experimentally and
numerically. We would like to point out that this situation is very
similar to many other examples in condensed-matter physics. The ubiquitous
Fermi-liquid fix point is strictly speaking always unstable because of
the Kohn-Luttinger effect\cite{ko.lu.65}. But for most metals this instability
occurs only at extremely low temperatures, and is practically irrelevant.
Another example is the ``ordinary'' superconductor to normal-state
transition at $T_c$. Strictly speaking, coupling to the
fluctuating electromagnetic field renders this fix point
unstable\cite{ha.lu.74}. 
However, this effect has never been observed
experimentally, since the associated critical region is too small.
Therefore, irrespective of the question of ultimate stability,
we argue that the \sof fix point is a robust one in a similar sense, and
it controls the physics near the AF and SC transitions.

\section*{Acknowledgments}

We would like to acknowledge useful discussions with
A. Aharony,
 E. Demler, X. Hu, and S. Sachdev.
This work is supported by the DFG 
via a Heisenberg
fellowship
(AR 324/3-1), 
by KONWIHR (OOPCV and CUHE),
as well as
by the NSF under grant numbers DMR-9814289.
The calculations were carried out at the high-performance computing centers 
HLRZ (J\"ulich) and LRZ (M\"unchen).

\ifnum\domybib>0
\bibliographystyle{prsty}
\bibliography{/users/arrigoni/references_database}
\else
\bibliographystyle{prsty}
\bibliography{references}
\fi

\end{document}